\documentstyle[aps,preprint]{revtex}
\begin{document}
\draft

\title{Langevin Dynamics of a Polymer with Internal
Distance Constraints}

\author{Michael P. Solf, Thomas A. Vilgis}

\address{Max-Planck-Institut f\"ur Polymerforschung,
Postfach 3148, 55021 Mainz, Germany}

\date{\today}

\maketitle

\begin{abstract}
We present a novel and rigorous approach to the Langevin
dynamics of ideal polymer chains subject to internal
distance constraints. The permanent constraints  are
modelled  by harmonic potentials in the limit when the
strength of the potential approaches infinity (hard
crosslinks).  The crosslinks are assumed to exist between
arbitrary pairs of monomers.  Formally exact expressions
for the resolvent and spectral density matrix of the system
are derived. To illustrate the method we study the
diffusional behavior of monomers in the vicinity of a
single crosslink within the framework of the Rouse model.
The same problem has been studied previously by Warner (J.
Phys. C: Solid State Phys.  {\bf 14}, 4985, (1981)) on the
basis of Lagrangian multipliers. Here we derive the full, hence
exact, solution to the problem.
\end{abstract}



\section{Introduction}

A theoretical treatment of the dynamics of polymer networks
is a generally unsolved problem. In a preliminary attempt
Edwards {\it et al.} \cite{edwa,deam} studied the problem
of a polymer subject to internal distance constraints. In
their investigation the underlying theoretical problem was
to handle the quenched degrees of freedom (hard crosslink
constraints) which, for example, in a random network exist
between pairs of arbitrary polymer segments (the monomers).
As a first step Edwards considered a (macroscopically)
long polymer chain which was internally crosslinked to
itself at random. The polymer backbone was assumed to be
Gaussian and the resulting dynamics was found to be of the
standard Rouseian type \cite{rouse,doied}. Permanent
junction points  were treated by Lagrangian multipliers,
which led to enormous technical difficulties for the
corresponding differential equations.  In fact, these could
only be handled by strong approximations, such as
pre-averaging in combination with harmonic variations.
Even when the problem was highly oversimplified and only
one crosslink was considered the method of Lagrangian
multipliers still becomes highly involved  as was pointed
out in a successive paper by Warner \cite{warner}.

The purpose of the present paper is to develop an
alternative  formalism for treating Langevin dynamics of
polymers subject to internal distance constraints. For
calculational simplicity the simplest working model for a
free polymer, the Rouse model \cite{rouse,doied}, is
considered. It is suggested that the more complicated
problem of a random network can also be treated {\em
exactly} by the presented method.  We adopt here the
minimal model suggested by Edwards \cite{edwa,deam} and
consider one (macroscopically) huge polymer chain which is
randomly crosslinked to itself. Such a crosslinking process
will lead to tetrafunctional crosslinks.  In previous works
we have already demonstrated that the analogous static
problem can be solved exactly when excluded volume effects
between the polymer segments are ignored
\cite{solvil,solf2}. Physical quantities such as the
static structure factor or the radius of gyration were
found to be self averaging and could be determined by
relatively simple numerical means.  The essential trick was
to account for the crosslinks in a general connectivity
matrix, that includes both the connectedness of the polymer
chain {\em and} an additional contribution from the
crosslinking. In close analogy we expect the corresponding
dynamic problem  to have a similar exact solution as long
as complicating factors such as excluded volume,
hydrodynamic forces or entanglements are neglected.  To
demonstrate this analogy we start from the standard
Langevin description for the polymer segments and solve the
stochastic differential equation in terms of its resolvent.
As an instructive example we reconsider the eight-shaped
polymer problem (i.e., a polymer ring with one
crosslink) studied by Warner \cite{warner} and
present its full solution. We first confirm the
results from Warner, which have been derived only
for low frequencies and low Rouse mode index, but show
secondly the exact solution in the entire frequency and
mode domain. Moreover the technique introduced here opens
new ways to study Langevin dynamics of constrained systems.

The paper is organized as follows. In section II the
physical model - a generalized version of the Rouse model
with internal distance constraints - is introduced. Section
III summarizes some of the basic theorems regarding
Langevin dynamics to be used later on.  In section IV - the
main calculational body of the paper - the general
mathematical formalism for handling internal distance
constraints is developed in detail. Our treatment is a
generalization of a method previously developed for
computing statistical properties of randomly crosslinked
Gaussian structures, i.e., ideal polymer networks
\cite{solvil,solf2}. In section V an application of the
method to diffusional motion of a single crosslink is given
(the Warner problem).  Section VI contains a short
discussion of main results and outlook.

\section{Rouse model with internal distance constraints}

As a minimal model for the dynamics of a Gaussian chain
subject to internal distance constraints we consider a
generalized version of the classical Rouse model
\cite{rouse}.  Its discrete version is a bead-spring model,
where the motion of the beads (monomers) is governed by the
coupled set of Langevin equations
\begin{equation}
\zeta \frac{d{\bf R}_i(t)}{dt}=-\nabla_{{\bf R}_i}{\cal H}_{0}
(\{ {\bf R}_i \})+ {\bf F}_i(t)~.
\end{equation}
In this equation of motion the inertial term is omitted as
usual. $\zeta$ denotes the inverse mobility or friction
constant, and ${\bf R}_i(t)$ ($i=0,...,N$) are the
trajectories of the monomers in 3-dimensional space. The
stochastic forces ${\bf F}_i(t)$ are assumed to be
$\delta$-correlated with first and second moments given by
\cite{doied}
\begin{eqnarray}
\langle F^\alpha_i(t) \rangle &=&0~, \\
\langle F_i^\alpha (t)F_j^\beta (t') \rangle &=&
2\zeta k_{\mbox{\tiny B}}T\delta_{ij}\delta_{\alpha\beta}
\delta (t-t')~.\nonumber
\end{eqnarray}
Superscripts $\alpha,\beta=x,y,z$ represent the 3-dimensional
Cartesian coordinates. In the classical Rouse model
excluded volume interaction and hydrodynamic forces are
disregarded and only elastic forces between monomers are
retained in the Hamiltonian. Here we consider a more
general form of the Rouse model with an extra potential to
allow for modelling the internal distance constraints
\begin{equation}
\beta {\cal H}_0=
\frac{3}{2a^2}\sum_{i=1}^N({\bf
R}_i-{\bf R}_{i-1})^2+\frac {3}{2\varepsilon
^2}\sum_{e=1}^M({\bf R}_{i_e}-{\bf R}_{j_e})^2~.
\end{equation}
The first term in the Hamiltonian represents the
connectivity of a Gaussian chain with persistence length
$a$, whereas the second term models the crosslinks. In
particular we are concerned with permanent constraints when
a monomer, say $i_1$, is linked to another monomer labeled
by $j_1$. For more than one crosslink a whole set C of
crosslink "coordinates" is needed to specify all junctions
in the system
\begin{equation}
{\mbox C}=(i_1,j_1),...,(i_e,j_e),...,(i_M,j_M)~.
\end{equation}
For example, depending on C the object under investigation
can be a flexible ring polymer, a two-dimensional membrane,
or a rubber network (figure 1).

Although the theory will be developed for arbitrary
coupling constant $\varepsilon$ two scenarios  are of
special relevance. For $\varepsilon \rightarrow 0$ it has
been shown \cite{solvil} that the Hamiltonian (3) is
suitable to model hard ${\delta}$-constraints (the
classical crosslinks) of the form
\begin{equation}
\prod_{e=1}^M\delta \big({\bf R}_{i_e}(t)-{\bf
R}_{j_e}(t)\big)~.
\end{equation}
The case $\varepsilon \rightarrow \infty$ leads to the
well-known problem of a free chain which serves here as a
reference state. One might be worried that the above model
is ill-defined and might diverge in the limit $\varepsilon
\rightarrow 0$.  It will be shown in section IV that the
converse is true  and that a surprisingly simple solution
can be obtained for this special limit. As shown
in the earlier paper on the static properties, ref. \cite{solvil}, it is
important to take the limit $\varepsilon \rightarrow 0$
at the very end of the calculaton. This procedure ensures firstly,
that no mathematical problems occure and secondly that in this case
hard crosslink constraints is treated properly.
Before going into
more of the calculational details some of the basic
definitions and notations regarding Langevin dynamics are
summarized in the next section.

\section{Langevin dynamics of ideal polymers: preliminaries}

Consider the generalized Ornstein-Uhlenbeck process
specified by equations (1)--(3). For calculational
simplicity matrix notation will be used. We define $N+1$
dimensional ``super-vectors" with three dimensional vector
components to account for the positions of all monomers
${\bf R}(t)=({\bf R}_0(t),..., {\bf R}_N(t))^\dagger$ and
for the stochastic forces acting upon them ${\bf
F}(t)=({\bf F}_0(t),..., {\bf F}_N(t))^\dagger$.  The
dagger denotes the conjugate complex of the transposed
vector.  Furthermore, the $N+1$ dimensional connectivity
(Kirchhoff) matrix is introduced as
\begin{equation}
{\cal M}(z)=\omega_0 \Big( {\cal W}_0+
\frac{1}{z} \sum_{e=1}^M {\cal X}(i_e,j_e) \Big)~,
\end{equation}
where
\begin{eqnarray}
{\cal W}_0&=&\left(
\begin{array}{rrrrr}
1 & -1 & 0 & \cdots & 0 \\ -1 & 2 & -1 &  & \vdots \\ 0 &
\ddots & \ddots & \ddots & 0\\
\vdots&  & -1 & 2 & -1 \\ 0 & \cdots & 0 & -1 & 1
\end{array}
\right ) ~~,
\end{eqnarray}
is the Wiener matrix associated with the polymer
``backbone", and
\begin{eqnarray}
{\cal X}(i_e,j_e)&=&\left(
\begin{array}{rrccccr}
0 & 0 & \cdots & 0 & \cdots & 0 & ~0 \\
\vdots & \vdots & & 0 & & \vdots &\vdots \\
0 & 1 & & \vdots & & -1 & 0 \\ 0 & 0 & \cdots & 0 & \cdots
& 0 & 0 \\ 0 & -1 & & \vdots & & 1 & 0 \\
\vdots & \vdots & & 0 & & \vdots &\vdots \\
0 & 0 & \cdots & 0 & \cdots & 0 & 0 \\
\end{array}
\right )
\begin{array}{c}
\vdots \\
\leftarrow i_e
\mbox{-th row} \\ \vdots \\
\leftarrow j_e
\mbox{-th row} \\ \vdots
\end{array} \nonumber \\
&~&
\end{eqnarray}
models a single crosslink. For further use we note that a
characteristic time scale is given by the inverse of the
``frequency"
\begin{equation}
\omega_0=\frac{3k_{\mbox{\tiny B}}T}{a^2\zeta}~.
\end{equation}
The dimensionless parameter $z= (\varepsilon /a)^2$ in
equation (6) is used to enforce the crosslinking
constraints.  With the above definitions the system of
stochastic differential equations (1) cast into matrix form
reads
\begin{equation}
\frac{d{\bf R}(t)}{dt}+{\cal M}(z)\, {\bf
R}(t)=\frac{1}{\zeta}\,{\bf F}(t)~.
\end{equation}
Some of the physical quantities of interest and their
interrelations are listed below. More  details can be
found for example in reference \cite{risken}. The Green's
function to equation (10) is given by
\begin{equation}
{\cal G}(t)=\lim_{z\rightarrow 0} e^{-{\cal M}(z)\,t}~.
\end{equation}
For $z\rightarrow 0$ the case of hard $\delta$-constraints
is recovered.  Otherwise  $z$ is an additional distance
parameter in the model. Of great importance in the
following derivation is the Laplace transform (resolvent)
of the matrix ${\cal M}$
\begin{equation}
{\cal R}(\omega)\equiv \int_0^\infty dt\,e^{-i\omega
t}\,{\cal G}(t)= \lim_{z\rightarrow 0}\, \Big(i\omega {\cal
I}+{\cal M}(z)\Big ) ^{-1}~,
\end{equation}
where ${\cal I}$ denotes the identity matrix.

>From equation (2) the spectral matrix of the Langevin
forces ${\bf F}(t)$ is found to be
\begin{equation}
\langle \tilde {\bf F}( \omega )  \tilde {\bf F}^\dagger(
\omega' ) \rangle = 12\pi k_{\mbox{\tiny B}}T \zeta\, \delta
(\omega-\omega' )\, {\cal I}~,
\end{equation}
where
\begin{equation}
\tilde {\bf F}( \omega )= \int_{-\infty}^\infty
dt\,e^{-i\omega t}\,{\bf F}(t)
\end{equation}
is the Fourier transform of the stochastic forces.  By
${\bf F}_i{\bf F}_j$ we mean the usual 3-dimensional scalar
vector product, whereas ${\bf F}{\bf F}^\dagger$ is used
for outer vector products.  Fourier transforms are  always
denoted by tilde.

A formal solution to equation (10) can  be obtained by
Rice's method \cite{risken}. The spectral density matrix
for the stochastic variable ${\bf R}(t)$ can  be derived by
use of (13) and Fourier transformation of (10)
\begin{equation}
\langle \tilde {\bf R}( \omega )  \tilde {\bf R}^\dagger(
\omega' ) \rangle = 12\pi D\, \delta (\omega-\omega' )\,
{\cal R}(\omega ){\cal R}^\dagger (\omega')~,
\end{equation}
with the diffusion coefficient $D$ given by
\begin{equation}
D=k_{\mbox{\tiny B}}T/\zeta~.
\end{equation}
Of primary interest for the diffusional behavior is the
two-time correlation function matrix defined as
\begin{equation}
{\cal C}(t,t')=\Big\langle \Big({\bf R}(t)-{\bf R}(t')
\Big)
\Big({\bf R}(t)-{\bf R}(t') \Big )^\dagger \Big\rangle~.
\end{equation}
Finally a steady-state solution for ${\cal C}(t,t')$ in
terms of the resolvent (12) is easily derived from the
expression for the spectral density matrix in (15)
\begin{equation}
{\cal C}(t-t')=\frac{12D}{\pi}\int_0^\infty d\omega\,
\Big (1-\cos \omega (t-t')\Big )\,{\cal R}(\omega ){\cal
R}(\omega )^\dagger .
\end{equation}

In the following study our primary goal will be to find a
general approach to calculate the resolvent ${\cal
R}(\omega )$, equation (12), for  an arbitrary set of
crosslinking constraints C, equation (4). From there
Green's function and correlation functions can in principle
be obtained by use of the standard formulas presented in
this section.  Although ${\cal M}(z)$ is a  matrix which
highly depends on all the details of C (the crosslink
positions), substantial progress can be made by invoking
the following exact method.

\section{Calculation of the resolvent ${\cal R}(\omega )$}

The first step in deriving a general expression for ${\cal
R}(\omega )$ for hard crosslinks is to find a way to
perform the limit $z\rightarrow 0$ in equation (12).  This
is an interesting problem in its own right which so far
could only been handled by introducing a finite cutoff at
$z=1$ and successive crude variational estimates. Here we
present an analytically exact approach that can overcome
these difficulties.  The mathematical trick is to utilize
an additional symmetry of the crosslink term in (8) by
writing the complete crosslink contribution in (6) in form
of a dyadic (outer vector) product
\begin{equation}
\sum_{e=1}^M {\cal X}(i_e,j_e) ={\cal U}\,{\cal U}^\dagger~,
\end{equation}
where
\begin{equation}
{\cal U}(\mbox{C})\equiv\big ( {\bf u}_1,...,{\bf u}_M\big
)
\end{equation}
has been introduced as the $(N+1)\times M$ rectangular
matrix with each of its $M$ column vectors given by
\begin{equation}
{\bf u}_e={\bf e}_{i_e} -{\bf e}_{j_e}~~~,~~(e=1,...,M).
\end{equation}
Here ${\bf e}_{i_e}$ represents the $N+1$ dimensional unit
vector with 1 in the $i_e$th position, and 0 otherwise.
Thus ${\cal U}(\mbox{C})$ has only $2M$ elements not equal
to zero that contain complete information about all
crosslink positions. In the above notation each crosslink
is uniquely represented by a vector ${\bf u}_e$. Note that all
vectors  ${\bf u}_e$, $e=1,...,M$ are linearly independent
for tetrafunctional crosslinks.  Combining equations (6),
(12) and (19)  the resolvent cast in matrix form reads
\begin{equation}
{\cal R}(\omega)= \frac{1}{\omega_0}\lim_{z\rightarrow 0}\,
\Big ( i\frac{\omega}{\omega_0}\, {\cal I}+{\cal
W}_0+\frac{1}{z}\, {\cal U}{\cal U}^\dagger \Big ) ^{-1}~.
\end{equation}
It is convenient to decompose (22)  into a singular and
nonsingular part with the nonsingular part being
\begin{equation}
{\cal W}=i\frac{\omega}{\omega_0} {\cal I}+{\cal W}_0~.
\end{equation}
It is well-known in the mathematical literature that if the
inverse of ${\cal W}$ exists, then the inverse in equation
(22) is given by
\begin{eqnarray}
\Big ({\cal W}&+&\frac{1}{z}\,{\cal U}{\cal U}^\dagger \Big
)^{-1}= {\cal W}^{-1}\\ &\times &
\Big ({\cal I}-{\cal U} \big (z{\cal I}+{\cal U}^\dagger
{\cal W}^{-1} {\cal U} \big)^{-1} {\cal U}^\dagger {\cal
W}^{-1} \Big )~.\nonumber
\end{eqnarray}
This theorem can be directly verified by matrix
multiplication.  The latter identity is also known as
Sherman-Morrison formula \cite{lancas}.

\subsection{The limit $z\rightarrow 0$}

There are  two subtle points about the existence of the
right hand side of (24). First we require ${\cal W}^{-1}$
to exist.  The only critical case arises if $\omega=0$,
i.e., when ${\cal W}={\cal W}_0$ in (23). The problem here
is that ${\cal W}_0$ is only positive {\it semi} definite
and there is one mode with eigenvalue 0 from translational
invariance.  This can be directly seen from the definition
of ${\cal W}_0$ in (7) which is a row (column) constant
matrix. However, even in the semi definite case the above
theorem remains valid if ${\cal W}^{-1}$ denotes a
generalized inverse of ${\cal W}$ as was proved in
reference \cite{lewis}.

Secondly from the definition of ${\bf u}_e$ in (21) it is
easily verified that for tetrafunctional crosslinks all $M$
vectors ${\bf u}_e$ are linearly independent. Thus in
general the kernel $z{\cal I}+{\cal U}^\dagger {\cal
W}^{-1} {\cal U}$ will be a {\it positive} definite matrix
of dimension $M$ and {\it full} rank which has only
positive eigenvalues  for all nonnegative values of $z$.
As a consequence performing the $z\rightarrow 0$ limit in
equation (24) leads to a well-defined expression for
the resolvent
\begin{equation}
{\cal R}(\omega)=\frac{1}{\omega_{0}} {\cal W}^{-1}\Big
({\cal I}- {\cal U} \big ({\cal U}^\dagger {\cal W}^{-1}
{\cal U}
\big)^{-1} {\cal U}^\dagger {\cal W}^{-1}\Big ) ~.
\end{equation}
The first term is the linear chain (Rouse) model, whereas
the second part arises entirely from the effect of
crosslinking. Although the case of general crosslinking
potential $z$ is still implicit in the basic formula (24),
we will restrict ourselves in the following discussion to
the somewhat simpler case $z=0$, i.e., hard
$\delta$-constraints.  Equations (24) and (25) are formally
exact solutions to the problem posed in equations (1)-(4).
The further evaluation of ${\cal R}(\omega )$ for specific
realizations of crosslinks C can be split into two parts
and is discussed in subsequent sections.

\subsection{Resolvent of ${\cal W}_{0}$}

Evaluation of the inverse of ${\cal W}$ in (25) can in
principle be done by full diagonalization of ${\cal W}_0$
which is tridiagonal. For calculational
simplicity we consider here only the cyclic counterpart of
${\cal W}_0$ with periodic boundary conditions
\begin{equation}
{\cal W}_0=\left(
\begin{array}{rrrrr}
2 & -1 & 0 & \cdots & -1 \\ -1 & 2 & -1 & \cdots & 0 \\
\vdots & \ddots & \ddots & \ddots & \vdots \\
0 & \cdots & -1 & 2 & -1 \\ -1 & \cdots & 0 & -1 & 2
\end{array}
\right ) ~.
\end{equation}
Both models (7) and (26) are known to obey the same Rouse
dynamics in the limit $N\rightarrow \infty$ \cite{zimm}.
Physically the latter situation represents a flexible ring
polymer. The eigensystem to (26) is of particular simple
form since it is a circulant.  The eigenvalues read
\begin{equation}
\lambda_k=4\sin^2 \frac{\pi k}{N+1}~~,~~k=0,...,N~.
\end{equation}
The modal matrix of (26) is the Fourier matrix ${\cal F}$
\cite{davis}  with matrix elements
\begin{equation}
[{\cal F}]_{kl}=\frac{1}{\sqrt{N+1}}\, \exp \frac{2\pi i k
l}{N+1}~~,~~ k,l=0,...,N~.
\end{equation}
Spectral decomposition leads to the well-known
representation of the inverse ${\cal W}^{-1}$ in terms of
its eigenvalues
\begin{equation}
[{\cal W}^{-1}]_{kl}=\frac{1}{N+1}
\sum_{n=0}^N  \frac{\exp \frac{2\pi i n(k-l)}{N+1}}
{i\omega/\omega_0+\lambda_n} ~.
\end{equation}

\subsection{Discussion of kernel}

The remaining calculational task for determining ${\cal
R}(\omega )$ is the evaluation of the kernel function in
the second part of (25)
\begin{equation}
{\cal K}(\omega;{\mbox C})\equiv ({\cal U}^\dagger {\cal
W}^{-1} {\cal U} \big)^{-1}~.
\end{equation}
Since ${\cal K}(\omega;{\mbox C})$ depends on all the
crosslink positions C=$(i_1,j_1),...,(i_M,j_M)$ via ${\cal
U}$ no further analytical progress is possible without
specifying the crosslink in the system. On the other hand,
from the mathematical structure of ${\cal K}(\omega;{\mbox
C})$ most problems of interest fall into one of the
following three categories. Only one of these will be
considered in detail in section V.

(i) The number of crosslinks $M$ is small. Since ${\cal
K}(\omega; {\mbox C})$ requires inversion of an $M\times M$
matrix analytical progress is always possible if $M$ is not
too large.  A particularly simple problem is treated in the
next section when we consider the dynamics of a polymer
shaped like the figure-of-eight (figure 1a).

(ii) Another special case arises when $M$ is large, but
there is some additional pattern in the structure of ${\cal
U}$.  Examples of this kind are illustrated in figures 1b
and 1c.  In particular the sketch in 1b shows an example of
a macromolecule with distance constraints $z\neq 0$, i.e.,
the more general case governed by equation (24). For the above
examples ${\cal K}$ can be calculated as a consequence of the
regularity of the crosslink positions. We will report on
these systems in a separate publication.

(iii) The third important category arises when $M$ is large
and the crosslink positions C in (4) are picked at random.
This is the case of a polymer gel (figure 1d).  Here one
has to resort to numerical computation of ${\cal K}(\omega;
{\mbox C})$ \cite{solf2}. However, there is still a huge
calculational advantage with (25). For a polymer network we
have in general $M\ll N$.
Equation (25) requires
``only" the inverse of an $M\times M$ matrix
\cite{solf2}
and not of the complete $N\times N$ connectivity matrix as
is commonly believed in the polymer literature \cite{eichinger,schulz}.
An analytic approach to the network problem would be to
perform the quenched average of the resolvent over the
crosslink positions C. The latter problem is a key problem
in current network research and has not been analytically
solved even for the static problem.\\

Before calculating ${\cal R}(\omega )$ for a specific
example, we want to establish some remarkable and general
properties of the operators in equation (25). Consider the
crosslink part in (25)
\begin{equation}
{\cal V}\equiv {\cal W}^{-1}{\cal U}\big ({\cal U}^\dagger
{\cal W}^{-1} {\cal U}
\big)^{-1} {\cal U}^\dagger {\cal W}^{-1}~.
\end{equation}
By elementary matrix multiplication it is found that
\begin{equation}
{\cal V}{\cal W}{\cal V}={\cal V}~,~ {\cal U}^\dagger {\cal
V}= {\cal U}^\dagger{\cal W}^{-1}~,~ {\cal V} {\cal
U}={\cal W}^{-1}{\cal U}~.
\end{equation}
A matrix with these properties is said to be a generalized
projector to ${\cal W}^{-1}$. Furthermore
\begin{equation}
({\cal W}{\cal V})^2={\cal W}{\cal V}~,~ ({\cal V}{\cal
W})^2={\cal V}{\cal W}
\end{equation}
are idempotents whose eigenvalues are known exactly:
$\lambda_1=1$ and $\lambda_2=0$ with degeneracies $M$ and
$N-M$. By use of the above results it is easy to prove  that
the resolvent  satisfies a remarkable orthogonality
relation
\begin{equation}
{\cal U}^\dagger {\cal R}(\omega )= {\cal R}(\omega
)\,{\cal U}=0~.
\end{equation}
Equation (34) is valid  for arbitrary crosslink positions C and
independent of the specific crosslink topology of the system under
investigation.

\section{Diffusional motion of a single crosslink}

As the simplest possible application of the method
developed in section IV we consider the figure-of-eight
shaped polymer depicted in  figure 1a.  What we have in
mind is to model the dynamics of a single crosslink in an
ideal dilute network when the distance between crosslinks
is large \cite{warner}. It is expected that monomers in
the neighborhood of the crosslink are somewhat affected by
the slower dynamics of the crosslink
\cite{edwa,warner,vilbo}.

A suitable realization of the system in figure 1a would be
\begin{equation}
{\cal U}(\mbox{C})={\bf u}_{1}={\bf e}_{0}-{\bf
e}_{\frac{N+1}{2}}~.
\end{equation}
That is, monomer 0 is linked to monomer $(N+1)/2$.  The
main calculational task is to determine the kernel function,
equation (30), of the system. From (35) and with ${\cal
W}^{-1}$ given by (29) we get immediately
\begin{equation}
{\cal K}(\omega )=\left (
\frac{4}{N+1}\sum_{n\,odd}
\frac{1}{i\omega /\omega_{0}+\lambda_{n}}
\right)^{-1}~,
\end{equation}
where the summation includes only the odd terms.  For the
diffusional motion the quantity of interest is the
self-correlation function contained in the diagonal
elements of the correlation matrix (18)
\begin{equation}
[{\cal C}(t-t')]_{ss}=\Big \langle
\big( {\bf R}_{s}(t)-{\bf R}_{s}(t')\big )^2 \Big \rangle~,
\end{equation}
where $s$ is the distance of the $s$th monomer with respect
to the crosslink at position $s=0$. Typical terms and
manipulations in the straight-forward derivation which is
not carried out in detail are of the form
\begin{eqnarray}
&&\frac{1}{N+1} \sum_{n\,odd}
\frac{\exp \frac{2\pi i s n}{N+1}}
{i\omega /\omega_{0}+\lambda_{n}} \nonumber \\ &\simeq&
\frac{1}{2}\int_{0}^{1} dx\, \frac{\exp (2\pi i s x)}
{i\omega /\omega_{0}+4\sin^2(\pi x)} \nonumber \\ &\simeq &
\frac{1}{4}
\sqrt{\frac{\omega_{0}}{\omega}}
\exp \left (-\frac{i\pi}{4}-(1+i)|s|
\sqrt{\frac{\omega}{2\omega_{0}}}\,
\right )~.
\end{eqnarray}
In deriving the first integral we have  performed the
$N\rightarrow \infty$ limit. The latter expression was
obtained by setting $\sin (\pi x)\simeq \pi x $. Only the
final result for the self-correlation function (37) is
quoted here
\begin{equation}
\Big \langle \big( {\bf R}_{s}(t)-{\bf R}_{s}(t')\big )^2
\Big \rangle =A(s,|t-t'|)\,2 a^2
\sqrt{\frac{\omega_{0}|t-t'|}{\pi}}~.
\end{equation}
The time-dependent prefactor is given by
\begin{eqnarray}
A(s,|t-t'|)&=&1-\frac{1}{2\sqrt{2\pi}}
\int_{0}^\infty dx\, \frac{1-\cos x}{x^{3/2}} \\
&\times & e^{-s'\sqrt{x}}
\Big (\cos\, (s'\sqrt{x})+\sin\, (s'\sqrt{x})
\Big ) \nonumber \\
&=&1-\frac{1}{4\pi} \int_0^\infty dy\, \cos \Big (
\frac{ys'^2}{2}-\frac{\pi}{4} \Big )\frac{\log (1+y^2)}
{y^{3/2}}~,
\end{eqnarray}
which scales with
\begin{equation}
s'\equiv \sqrt{\frac{2s^2}{\omega_{0}|t-t'|}}~.
\end{equation}
The complicated integral in (41) is plotted in figure 2.
The asymptotic behavior for small values of $s'$ is
governed by the expansion
\begin{equation}
A(s')=\frac{1}{2}+\frac{s'\sqrt{\pi}}{2\sqrt{2}}-\frac{s'^2}{4}
+{\cal O}(s'^3)~.
\end{equation}
In particular for the diffusional motion of the crosslink
$(s=0)$, we find $A=1/2$ which is exactly half the
diffusion constant of an unconstrained monomer in the Rouse
model \cite{degen}. The finding is in agreement with the
result in reference \cite{warner} based on the method of
Lagrangian multipliers. For $s\rightarrow\infty$ (monomers
that are sufficiently far from the crosslink) we recover
the diffusion law of the classical Rouse model ($A=1$)
which was first derived by de Gennes \cite{degen}.

In addition we obtain the crossover from the slower
dynamics of the crosslink to that of a ``free" monomer in
the classical Rouse model as $s$ is varied from zero to
infinity.  The crossover takes place on time scales of the
order
\begin{equation}
\tau_s=s^2/\omega_0=(s a)^2/(3 D)~,
\end{equation}
where $sa$ measures the distance of the monomer from the
crosslink (figure 3).  Interestingly a monomer begins to
feel the presence of the crosslink only after a timespan
of the order $\tau_s$.

The two limiting cases $A=1/2$ relevant for the slower
dynamics of the crosslink and $A=1$  for the ``free" chain
segments far away from the crosslink are  expected on
physical grounds (dashed lines in figure 3). An ``inner"
chain segment has only two neighbors, whereas the crosslink
is surrounded by four neighbors. Thus in general for a
monomer with functionality $f$ a prefactor  $A(f) = f/2$ is
expected as was pointed out previously \cite{warner}.

\section{Conclusion}

Within the framework of the Rouse model we have proofed
that an exact solution for the Langevin dynamics of a
polymer subject to {\em hard} delta constraints exists when
excluded volume and hydrodynamic forces are neglected.  The
fundamental and general result for the resolvent, equation
(25), was derived for an arbitrary crosslink configuration
making our results also applicable to the challenging
problem of a random network.

In this investigation  we restricted ourselves to the
simplest physical scenario, where only Rouseian dynamics
was involved. This case was deliberately chosen to
highlight the principal mathematical difficulties. As a
special application we studied the dynamics of the
figure-of-eight shaped polymer depicted in figure 1a. In
contrast to an earlier attempt by Warner \cite{warner}
based on Lagrangian multipliers which yielded only two
limiting cases $s\rightarrow 0$ and $\infty$, the full
solution could be derived by our method.  Moreover, our
result allows for computation of the dynamic scattering
function \cite{degen} and comparison with experimental data
taken in the dry network state. A detailled
comparison will be studied in a
future, less formal paper.

For the physically more realistic scenario of a swollen
network in a Theta solvent further generalizations are
required like taking hydrodynamic interaction into account.
A generalized version of the equation of motion (1) would
read
\begin{equation}
\zeta \frac{d{\bf R}_i(t)}{dt}=
\sum_{j} {\bf \Theta}_{i,j}
\left( -\nabla_{{\bf R}_j}{\cal H}_{0}
(\{ {\bf R}_j \})+ {\bf F}_j(t) \right)~,
\end{equation}
where ${\bf \Theta}_{i,j}$ is the Oseen tensor.  Although
the above equation becomes analytically untractable, for
most experimental situations a pre-averaged treatment is
well justified \cite{doied}. This computation is left for
future work.


\acknowledgements
MPS wishes to acknowledge financial support from the
Deutsche Forschungs\-gemeinschaft,
Sonder\-forschungs\-bereich 262.


\begin{figure}
\caption[1]{Examples of different crosslinking topologies.
(a) The polymer shape discussed in section V. $a$ is
the persistence length of the polymer backbone. The hard
crosslink constraint is enforced by $\varepsilon
\rightarrow 0$. (b) A ladder shaped polymer with
$\varepsilon \neq 0$. (c) Two-dimensional membrane.
(d) Random network.}
\label{fig1}
\end{figure}

\begin{figure}
\caption[2]{Prefactor $A(s')$, equation (41).
The dashed lines represent the asymptotic behavior for
small and large values of $s'$.}
\label{fig2}
\end{figure}

\begin{figure}
\caption[3]{Crossover of the mean squared displacement from
the dynamics of a ``free" Rouseian monomer (upper dashed
line) to the slower dynamics of a crosslink (lower dashed
line).}
\label{fig3}
\end{figure}


\begin{references}

\bibitem{edwa} S. F. Edwards, J. Phys. A: Math. Gen. {\bf
7}, 318 (1974).

\bibitem{deam} S. F. Edwards in {\it Polymer Networks},
eds.  A. J. Chompff and S.  Newman (Plenum Press, New York,
1971); R. T. Deam and S. F. Edwards, Proc.  Trans. R. Soc.
London A {\bf 280} (1976) 317; R. C. Ball and S. F.
Edwards, Macromolecules {\bf 13} (1980) 748.

\bibitem{rouse} P. E. Rouse, J. Chem. Phys. {\bf 21}, 1272
(1953).

\bibitem{doied} M. Doi and S. F. Edwards, {\it The Theory
of Polymer Dynamics} (Clarendon Press, Oxford, 1986), chap.
4.

\bibitem{warner} M. Warner, J. Phys. C: Solid State Phys.
{\bf 14}, 4985 (1981).

\bibitem{solvil} M. P. Solf and T. A. Vilgis, J. Phys. A:
Math. Gen. {\bf 28}, 6655 (1995).

\bibitem{solf2} M. P. Solf and T. A. Vilgis,
J. Phys. France I, submitted.


\bibitem{risken} H. Risken, {\it The Fokker-Planck
Equation} (Springer-Verlag, Berlin, 1989), chap. 3.

\bibitem{lancas} P. Lancaster and M. Tismenetsky, {\it
Theory of Matrices} (Academic Press, San Diego, 1985).

\bibitem{lewis} T. O. Lewis and T. G. Newman, SIAM J. Appl.
Math.  {\bf 16}, 701 (1968).

\bibitem{zimm} B. H. Zimm and R. W. Kilb, J. Polymer Sci.
{\bf 37}, 19 (1959).

\bibitem{eichinger} B.E. Eichinger, J.E. Martin,
J. Chem. Phys. {\bf 69}, 4595 (1978)

\bibitem{schulz} M. Schulz, P. Reineker, M. M\"oller,
J. Chem. Phys. {\bf 103}, 10701 (1995)

\bibitem{davis} P. J. Davis, {\it Circulant Matrices} (Wiley,
New York, 1979).


\bibitem{vilbo} T. A. Vilgis and F. Bou\'{e}, J. Polymer
Sci.: Part B {\bf 26}, 2291 (1988).

\bibitem{degen} P. G. de Gennes, Physics {\bf 3}, 37
(1967).
\end{references}
\end{document}